\documentclass[PRApplied, twocolumn, amsmath,amssymb, aps]{revtex4-1} 
\usepackage{graphicx} 
\usepackage{hyperref} 
\usepackage{color}
\usepackage[yyyymmdd,hhmmss]{datetime}
\usepackage[normalem]{ulem}
\usepackage{todonotes}
\usepackage{multirow}
\usepackage{float}
\usepackage{array}
\usepackage{adjustbox}
\usepackage{makecell}
\usepackage{longtable}
\usepackage{booktabs}

\begin{document}
\title{Enhanced spin transfer torque in platinum/ferromagnetic-metal structures by optimizing the platinum thickness} 
\author{Jianshu Xue} 
\affiliation{School of Physics, State Key Laboratory of Crystal Materials, Shandong University, 27 Shandanan Road, Jinan, 250100 China}
\author{Yaping Guo} 
\affiliation{School of Physics, State Key Laboratory of Crystal Materials, Shandong University, 27 Shandanan Road, Jinan, 250100 China}
\author{Ledong Wang} 
\affiliation{School of Physics, State Key Laboratory of Crystal Materials, Shandong University, 27 Shandanan Road, Jinan, 250100 China} 
\author{Yanan  Dong} 
\affiliation{School of Physics, State Key Laboratory of Crystal Materials, Shandong University, 27 Shandanan Road, Jinan, 250100 China}
\author{Yanxue Chen} 
\affiliation{School of Physics, State Key Laboratory of Crystal Materials, Shandong University, 27 Shandanan Road, Jinan, 250100 China}
\author{Yufeng Tian} 
\affiliation{School of Physics, State Key Laboratory of Crystal Materials, Shandong University, 27 Shandanan Road, Jinan, 250100 China}
\author{Shishen Yan} 
\affiliation{School of Physics, State Key Laboratory of Crystal Materials, Shandong University, 27 Shandanan Road, Jinan, 250100 China}
\author{Lihui Bai} 
\email{lhbai@sdu.edu.cn}
\affiliation{School of Physics, State Key Laboratory of Crystal Materials, Shandong University, 27 Shandanan Road, Jinan, 250100 China} 
\begin{abstract} 
Spin transfer torque (STT) driven by a charge current plays a key role in magnetization switching in heavy-metal/ferromagnetic-metal structures. 
The STT efficiency defined by the ratio between the effective field due to STT and the current density,  is required to be improved to reduce energy compulsions in the STT-based spintronic devices.
In this work, using the harmonic Hall measurement method, we experimentally studied the STT efficiency in platinum(Pt)/FM structures as a function of the Pt thickness.
We found that the STT efficiency strongly depends on the Pt thickness and reaches a maximum value of 4.259 mT/($10^6$A/$cm^{2}$) for the 1.8-nm-thickness Pt sample.
This result indicates that competition between spin Hall effect (SHE) and Rashba effect as well as spin diffusion process across the Pt layer determines the Pt thickness for the maximum STT efficiency. 
We demonstrated the role played by the spin diffusion besides the spin current generation mechanisms in improvement of the STT efficiency, which is helpful in designing STT-based devices.
\end{abstract}

\maketitle
\section{introduction}
Spin transfer torque(STT) is that a spin current exerts a torque on the magnetization due to exchange interaction, which was demonstrated as an viable way to realize power-efficient magnetization switching compared to that by an Oersted field\cite{Liu2010apl, Liu2012PRL, Liu2012science, Nguyen2015apl}. 
For the first generation of STT, the spin current was polarized by a ferromagnetic metallic layer in spin valves\cite{Bedau2010apl} or magnetic tunneling junctions\cite{JackNP2008, ChungNP2009}. 
Later, a pure spin current was produced via spin Hall effect in heavy metals (SHE)\cite{Liu2012science} and via Rashba effect at an interface with a structure-inverse-asymmetry\cite{huan2014scientific} to exert a STT on a magnetization which is also known as spin-orbit torque(SOT).
Along this procedure of charge current leading to a spin current acts on a magnetization as a torque, the STT efficiency is defined as the ratio between the effective field $\Delta H$ on the magnetization due to the STT and the charge current density $j_{c}$ applied to the heavy metalic layer\cite{MC2010APL}.
The effective field $\Delta H$ due to STT has two components, along the current direction (longitudinal component) and along the direction perpendicular to the current and the magnetization (transverse component), which corresponding to the in-plane STT and the perpendicular STT\cite{BergerPRB1996, SlonczewskiMMM1996}. 
These effective fields are linearly proportional to the amplitude of the spin current density $j_{s}$ arrived the magnetization, therefore, the STT efficiency  is proportional to the ratio between the spin current density $j_{s}$ and the charge current density $j_{c}$.
 Improving the STT efficiency lies at the center of spin transport properties and applications of STT in spintronics devices.

To improve the STT efficiency, many works have been done mainly in two scenarios.
One is choosing materials with strong spin-orbit interaction to enhance the conversion from a charge current to a spin current.
Heavy metals were explored  to evaluate the spin Hall angle used in FM/heavy metal junctions, such as Platinum($\theta_{SH}^{Pt} \sim$ 0.08 \cite{NVlietstraprb2013}) ,  Tantalum($\theta_{SH}^{Ta} \sim$ 0.12 to 0.15 \cite{Liu2012science}) and PtAl alloy \cite{Minh2016apl}) and so on.
Another attempt is combining two or more heavy metals to further enhance the spin current conversion.
For example,  Pt/W/CoFeB\cite{  Ma2018PRL} and Pt/CoNi/W \cite{Yu2016apl} were designed to enhance the STT efficiency.
Even though, a high charge current density is still necessary in magnetization switching in ferromagnetic materials. 
To further enhance the STT efficiency, new solution is required in device designing, which could be done by optimizing the spin diffusion process besides by improving the charge-spin current conversion. 

In this work, we enhance the STT efficiency by optimizing the thickness of the heavy metal layer, which is rooted into the cooperation between the bulk effect(SHE) and the interface effect (Rashba effect) as well as spin diffusion process.  
We experimentally demonstrated that the STT efficiency in Ta/Pt/FM structures can be enhanced while the Pt layer has a specific thickness.  
The enhanced STT efficiency is almost 4 times larger than those reported\cite{Ma2018PRL, MustafaAkyol2015apl, Kong2018APL, JaeWookLee2017PRB, Yu2017ad, P2017MMM}. 
Our work provides a new approach to further reduce power consumption in magnetization switching. 

\section{Charge-Spin conversion and spin diffusion}
\begin{figure}[htb]
    \includegraphics[width = 8.17 cm]{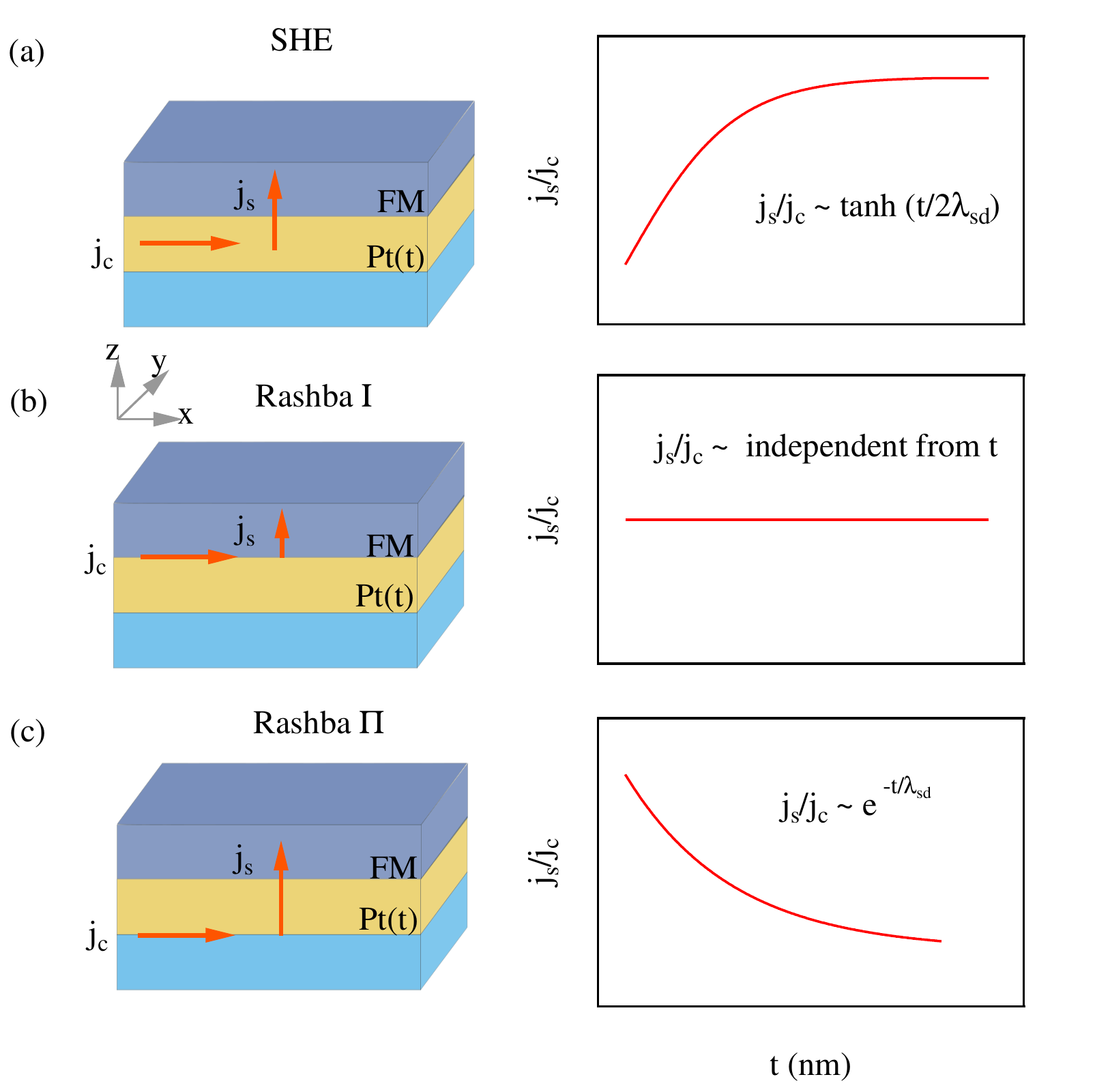}
    \caption{(a) A sketch of SHE. SHE in Pt is a bulk effect, spin current is produced by the charge current in Pt layer due to SHE and diffuses to the adjacent FM layer. $j_{s}/j_{c}$ related to tanh(t/$\lambda_{2sd}$) is limited by spin diffusion mechanism, where t is the thickness of Pt, $\lambda_{sd}$ is the Pt diffusion length. 
(b) The Rashba effect at the FM layer and the bottom Pt layer interface. Spin polarization induced by interfacial electrical field diffuses into the adjacent ferromagnetic layer formed spin current, and exerts the spin torque on the FM magnetization. The magnitude of $j_{s}/j_{c}$ is independent on Pt thickness. 
(c) The Rashba effect from the interface of the bottom Pt layer and the buffer layer Ta. Due to spin dissipation in Pt layer, the magnitude of $j_{s}/j_{c}$ is relate to $e^{-t / \lambda_{sd}}$. }
    \label{fig1}
\end{figure}

The STT efficiency depends on the spin current density $j_{s}$ arrived the magnetization for a given charge current density $j_{c}$ in the heavy metallic layer besides the spin diffusion in the FM layer. 
Therefore, the spin diffusion process is as important as the charge-spin conversion process. 
As shown in Fig. \ref{fig1}(a), a charge current density $j_{c}$ in the Pt layer produces a spin current density $j_{s}$ via SHE and the spin current diffuses through the heavy metallic layer (here it is Pt) and injects into the adjacent ferromagnetic layer.
The STT efficiency is proportional to $j_{s}/j_{c}$, which is determined by the spin Hall angle $\theta_{SH}$ and spin diffusion length $\lambda_{sd}$ in the Pt layer as $j_{s}/j_{c} \propto \theta_{SH} tanh (t/2\lambda_{sd})$\cite{O2010PRB}. 
Here, t is the thickness of the Pt layer. 
There are two interfaces aside the Pt layer with two spin current sources due to Rashba effect as shown in Fig. \ref{fig1}(b) and (c).
When the charge current flows along the interface between FM and Pt, the STT efficiency is $\propto j_{s}/j_{c} \propto R_{1}$, which is independent on the thickness of the Pt layer.
Here $R_{1}$ is the Rashba parameter at the Pt/FM interface depending on the chemical potential asymmetry along the normal direction\cite{JC2013naturec}.
While the charge current flows along another interface of the Pt layer as shown in Fig. \ref{fig1}(c), the produced spin current by the Rashba effect diffuses across the Pt layer and reaches the FM layer.
Thus, the STT efficiency is $\propto j_{s}/j_{c} \propto R_{2} e^{-t/ \lambda_{sd}}$.
Here $R_{2}$ is the Rashba parameter at the second Pt interface away from FM layer.
Additionally, other spin current source (such as SHE in Ta layer\cite{Liu2012science, Kim2013nmat}) at the same side of the Pt layer would roughly follows the same diffusion process.  

Taking all three types of spin currents and their diffusion processes into account, we will find that the STT efficiency depends on the thickness of the Pt layer, 
\begin{equation}
\frac{\Delta H}{j_{c}} = Amp (\theta_{SH} tanh (t/2\lambda_{sd}) + R_{1} + R_{2} e^{-t/ \lambda_{sd}})  
\label{equation1}
\end{equation}
Here, $\theta_{SH}$ = $j_{s,SHE}^{0}$ / $j_{c}$, $R_{1}$ = $j_{s,R_{1}}^{0}$ / $j_{c}$, $R_{2}$ = $j_{s,R_{2}}^{0}$ / $j_{c}$.
$j_{s,SHE}^{0}$, $j_{s,R_{1}}^{0}$, $j_{s,R_{2}}^{0}$ are the local spin currents generated by SHE, Rashba effect and Rashba type effect at the second interface.
Amp is related to spin mixing conductance, spin diffusion in FM layer and magnetization orientation, but is independent from the Pt thickness t.
The first term indicates that the STT efficiency increases gradually with increasing the Pt thickness and tends to be saturated eventually, the second term is a constant term which is independent on the Pt thickness, and the third term decreases exponentially with the increase of Pt thickness t,  as illustrated in Fig.\ref{fig1}(a), (b) and (c) respectively.

\section{Samples and EXPERIMENTAL METHODS}
\begin{figure}[htbp]
    \includegraphics[width =8.17 cm]{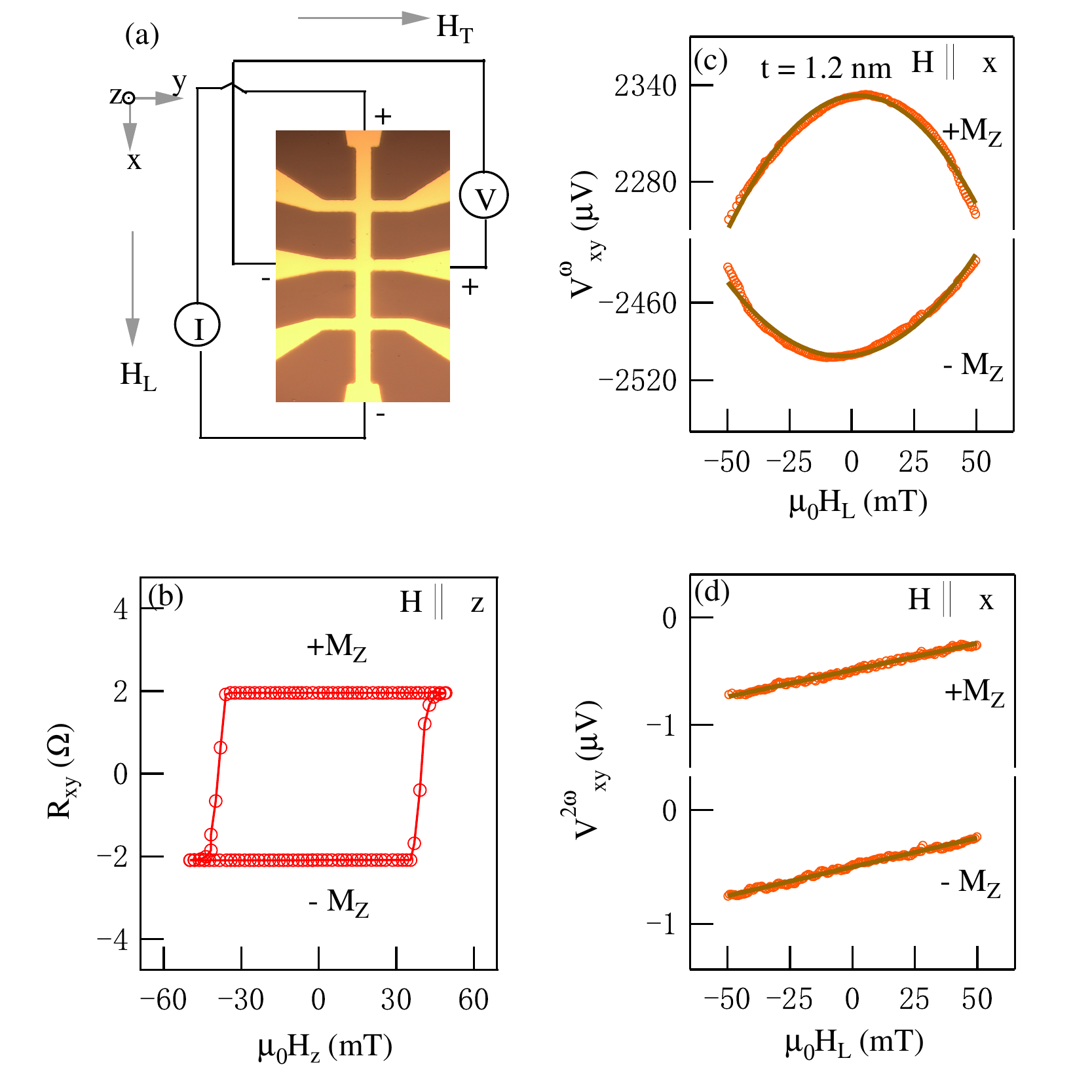}
    \caption{(a) Optical micrograph of the patterned Hall bar and the electrical measurement set up. AC current is applied along x axis, and the voltages of $V_{xy}^{\omega}$ and $V_{xy}^{2\omega}$ are measured perpendicular to the current direction with the in-plane longitudinal $H_L$. (b) The anomalous Hall resistance ($R_{xy}=V_{xy}/I$) dependent on the out-of-plane magnetization $H_Z$. (c) First and (d) second harmonic Hall voltages with longitudinal swept field $H_{L}$ for both magnetization along +z axis ($+M_Z$) and magnetization along -z axis ($-M_Z$) at the Pt thickness t = 1.2 nm. }
    \label{fig2}
\end{figure}

We fabricated Pt/FM devices to evaluate the STT efficiency dependent on the Pt thickness.
Before depositing the Pt/FM bilayer, a 2-nm-thick Ta layer was first sputtered on the thermally oxidized silicon substrates as a buffer layer to improve the quality of the metallic layer on top.
And, this Ta buffer layer introduces the second interface to the Pt/FM layer.
The Pt layer was deposited using magnetron sputtering with a base pressure of 3 $\times$ $10^{-8}$ mTorr  at room temperature.
The thickness t of the Pt varied from 1 nm to 7 nm to allow us to compare the effective field due to STT in different samples.
The FM layer is a $[Co(0.6)/Pt(0.6)]_4$ super lattice with an additional Pt capping layer for protection.
Then the thin films were patterned into Hall bars of 200 $\mu m$ long and 10 $\mu m$ wide as shown in Fig. \ref {fig2} (a) using photolithography and Ar ion milling technique. 
The $[Co(0.6)/Pt(0.6)]_4$ super lattice exhibited strong perpendicular magnetic anisotropy (PMA) in the Hall measurement as shown in Fig. \ref{fig2}(b). 
In the Hall measurement, a charge current of 1.5 mA was applied along the x axis and an external magnetic field  $H_{z}$ was aligned along the normal direction.
The square shape in the Hall resistance was attributed to the anomalous Hall effect induced by the magnetization with PMA. 
Thus, we are allowed to evaluate the STT efficiency using harmonic Hall measurment\cite {Hibino2017apl, MustafaAkyol2015apl, SeonghoonWoo2014apl, KHasegawa2018PRB, Li2017APL, Kong2018APL, Krishnia2019MMM, Zhang2016PRB, JaeWookLee2017PRB, Pai12014APL, SatoruEmori2014APL, li2017sc, Yu2017ad, P2017MMM}.

In the harmonic Hall measurement as shown in Fig. \ref{fig2}(a), an amplitude modulated current was applied to the Hall bar along the x axis and we measured the first harmonic Hall voltage $V_{xy}^{\omega}$ and the second harmonic Hall voltage $V_{xy}^{2\omega}$ using lock-in technique.
Different from a normal Hall measurement, the external magnetic field was applied in plane of the thin film, along x axis  and along y axis known as longitudinal field $H_{L}$ and transverse field $H_{T}$ respectively. 
 Due to PMA magnetization in the FM layer, we know that  the first harmonic Hall voltage $V_{xy}^{\omega}$ indicates the magnetization projection is directed along the normal direction while the external magnetic field forces the magnetic moment to deviate from z axis. 
Therefore, the $V_{xy}^{\omega}$ makes a parabolic curve as a function of the external magnetic field.
Following Kim's work\cite{Kim2013nmat}, the second harmonic Hall voltage $V_{xy}^{2\omega}$ indicates the strength of the effective field due to STT change gradually. 
Thus, the effective field due to STT can be evaluated by measuring both $V_{xy}^{\omega}$ and $V_{xy}^{2\omega}$ using $\Delta H_{L (T)}= {-2} \frac {\partial V^{2 \omega}_{xy}} {\partial H_{L (T)}} / \frac {\partial ^{2} V^{\omega}_{xy}} {\partial H ^{2}_{L (T)}}$.
Here, subscript L and T denote the longitudinal field and the transverse field which are measured respectively.
As shown in Fig. \ref{fig2}(c) and (d),  we obtain the $\Delta H_{L}$ in this sample is about 1.6 mT, which is five times larger than the $\Delta H_{T}$ $\approx$ -0.3 mT (raw data  not shown here).
Taking the current density of  $cm^{2}$ into account, we get the STT efficiency which are 3.1 mT/($10^6$A/$cm^{2}$) along the x axis and  -0.58 mT/($10^6$A/$cm^{2}$) along y axis in the 1.2-nm-thick-Pt sample.  
We found that the longitudinal effective field due to STT is an order of magnitude larger than the transverse effective field due to STT in these serial samples. 
And the longitudinal effective field corresponds to the in-plane STT torque. 
All the sample were measured in the identical condition at room temperature. 
 
\section{RESULTS AND DISCUSSION}
\begin{figure}[htbp]
    \includegraphics[width =8.17 cm]{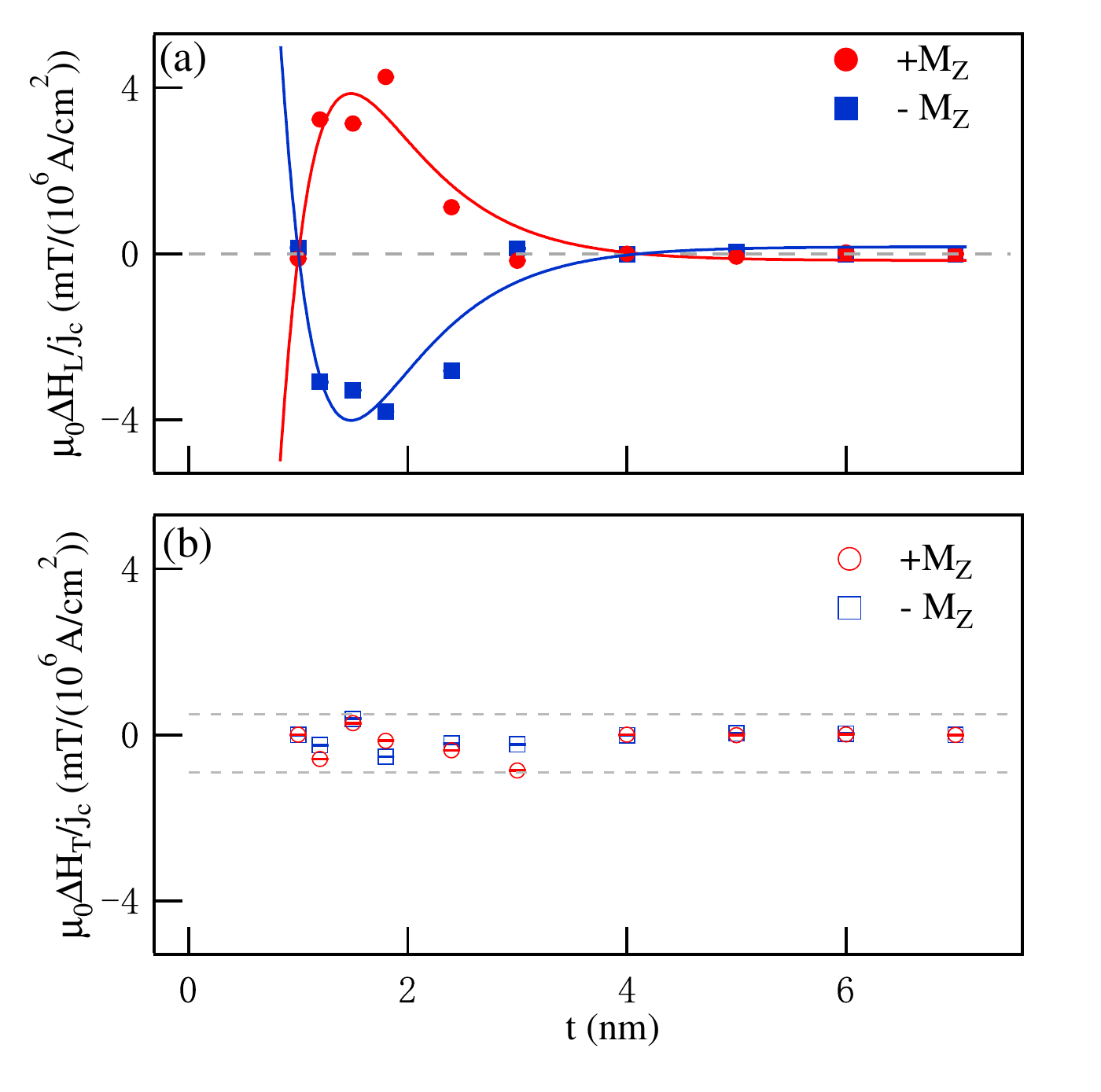}
    \caption{(a) $\Delta H_{L}/j_{c}$ and (b) $\Delta H_{T}/j_{c}$ plotted as a function of the bottom Pt thickness t. The red circles represent $\Delta H/j_{c}$ with $+M_Z$ and the blue squares represent $\Delta H/j_{c}$ with $-M_Z$. The solid line indicates the fitting results according to Eq. (\ref {equation1}).}
    \label{fig3}
\end{figure}

Fig. \ref{fig3} summarized the STT efficiency as a function of the thickness t of the Pt layer for both longitudinal and transverses effective field respectively.
Here, the solid circles represent the magnetization along z axis ($+M_Z$) and the solid squares represent the magnetization along -z axis ($-M_Z$). 
The STT efficiency has a large longitudinal component $\Delta H_{L}$ while the transverse component $\Delta H_{T}$ is almost ignorable.
It is clear that the $\Delta H_{L}$ increases by decreasing the Pt thickness from 7 nm and reaches a maximum around 1.8 nm, then decreases down to zero and crosses zero when the thickness t is less than 1.1 nm. 
Such a non-monotonic behavior could not be explained only using the SHE and spin diffusion in the Pt/FM bilayer.
Considering Rashba effect at both interfaces of Pt and the SHE in Pt, we fitted the $\Delta H_{L}$ using Eq. (\ref{equation1}).
Here, spin Hall angle $\theta_{SH}$ in the Pt layer is adapted of 0.08\cite{NVlietstraprb2013} which is quite close to the Pt property in this devices\cite{Wang2019PRApplied}. 
With this  $\theta_{SH}$ = 0.08, the fitting allows us to determine both the Rashba effect at the Pt/FM interface $R_{1}$ and the charge-spin current conversion efficiency at Pt/Ta interface due to Rashba effect and the SHE in Ta $R_{2}$. 
An additional parameter Amp was used as a free parameter to justify the amplitude, which includes the spin mixing conductance at the FM/Pt interface and spin diffusion process in the limited thickness of the FM layer.
The STT efficiency for magnetization along both z and -z axis were fitted simultaneously using the same parameters of $R_{1}$, $R_{2}$ and spin diffusion length $\lambda_{sd}$.
The fitting results were highlighted by the solid curves in Fig. \ref{fig3}(a) with parameters of $R_{1}$=-0.081, $R_{2}$=0.129, $\lambda_{sd}$=0.676 nm and Amp of $\pm$ 2850 $mT/(10^{6}A/cm^{2})$(sign depending on the magnetization orientation).
The results indicate that the Rashba effect $R_{1}$ at the interface of FM/Pt has a comparable charge-spin conversion efficiency with spin hall angle but with an opposite sign.
$R_{2}$ is larger than the spin Hall angle in Pt layer, which is an effective Rashba effect covering the Rashba effect and SHE in Ta layer. 
The competition between the Rashba effect and the SHE makes the sign of STT efficiency change at around t = 1.1 nm as shown in Fig. \ref{fig3}(a).
Eq. (\ref{equation1}) predicts that a thin Pt layer (less than 0.8 nm) has a large STT efficiency. 
However, the Pt layer thickness is limited by the roughness of the thin film in this work.  
Both experimental data and theoretical prediction show that a 1.8-nm-thickness Pt layer leads to a maximum of STT efficiency with a large longitudinal effective field.

\begin{figure}[htbp]
    \includegraphics[width =8.17 cm]{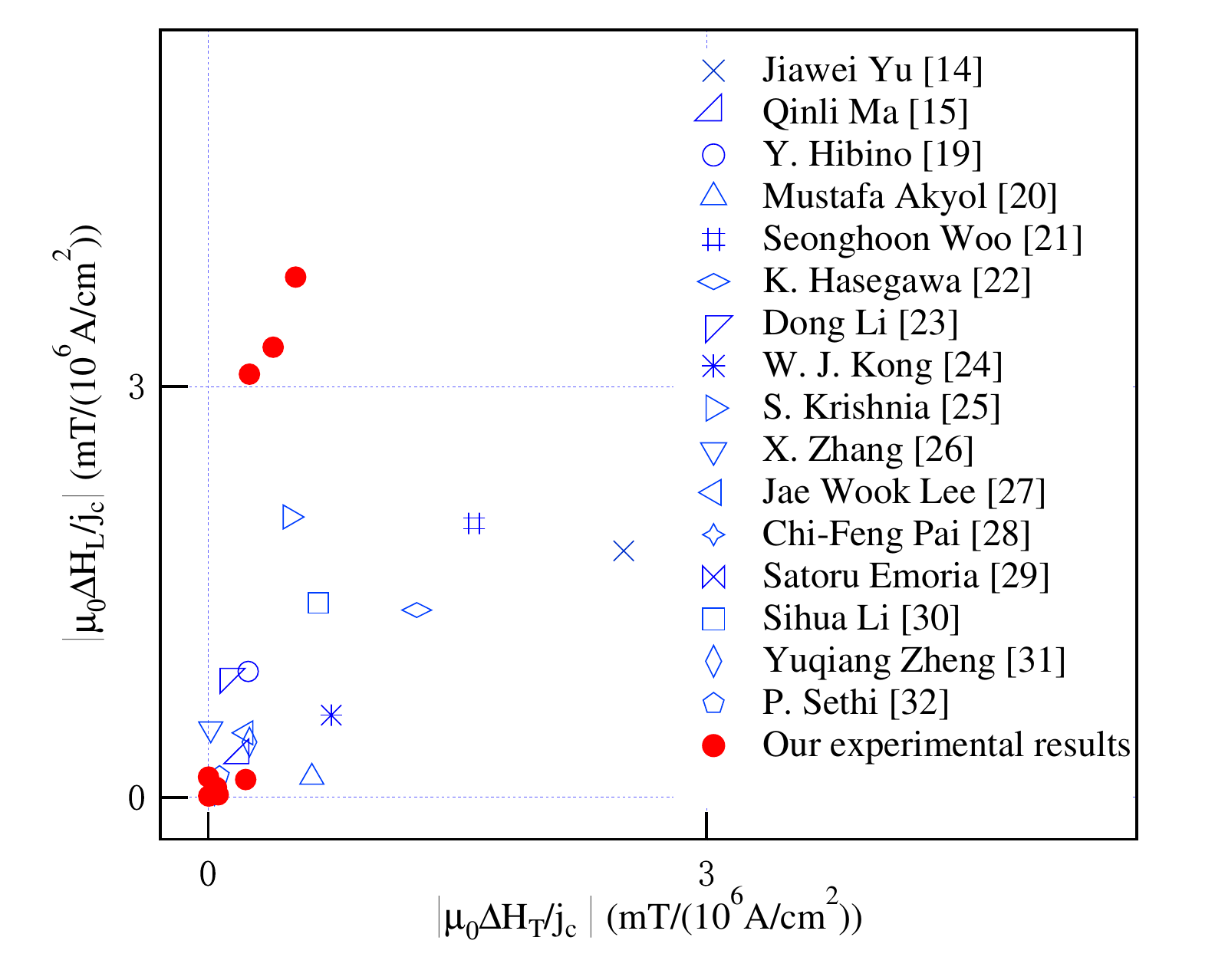}
    \caption{The solid red circles represent the ratio of $\Delta H_{L}/j_{c}$ and $\Delta H_{T}/j_{c}$ with different Pt thickness t in our experimental results. The blue symbols represent data from the references for comparison. }
    \label{fig4}
\end{figure}

 The STT efficiency depending on the thickness of the heavy metallic layer thickness was also been observed in similar structures, such as Pt/W/CoFeB and Pt/CoNi/W\cite{  Ma2018PRL, Yu2016apl}. 
 Those results can also be well fitted with Eq. (\ref{equation1}) with the reasonable parameters, hints that the spin diffusion in the heavy metallic layer plays an important role in improvement of the STT efficiency.

In Fig. \ref{fig4}, such an enhanced STT efficiency was achieved by optimizing the Pt thickness compared to the reported results\cite{Yu2016apl, Ma2018PRL, Hibino2017apl, MustafaAkyol2015apl, SeonghoonWoo2014apl, KHasegawa2018PRB, Li2017APL, Kong2018APL, Krishnia2019MMM, Zhang2016PRB, JaeWookLee2017PRB, Pai12014APL, SatoruEmori2014APL, li2017sc, Yu2017ad, P2017MMM}.
Both axis are the transverse component and longitudinal component of the STT efficiency. 
The open symbols summarized the reported results and the solid circles are from this work.
It is clear that the STT efficiency from most groups crowed around the 1 mT/($10^6$A/$cm^{2}$).
Our maximum value reaches 4.259 mT/($10^6$A/$cm^{2}$) especially along the longitudinal component, which is almost 4 times larger than the reported results\cite{Ma2018PRL, MustafaAkyol2015apl, Kong2018APL, JaeWookLee2017PRB, Yu2017ad, P2017MMM}.
The STT efficiency of the transverse component is still weak that is might related to the spin diffusion process in the FM layer\cite{zhang2014PRL, zhang2004PRL}.

 \section{Summary}
In this work, using the harmonic Hall measurement, we experimentally studied the STT efficiency in Pt/FM multi-layer structures as a function of the Pt thickness.
We found that the STT efficiency strongly depends on the Pt thickness and reaches a maximum for a 1.8-nm-thickness Pt sample.
We analyzed the spin diffusion process in Pt/FM multi-layer structures and explained the observation by taking the SHE in Pt layer and Rashba effect at interfaces into account. 
Competition of SHE and Rashba effect and spin diffusion process across the Pt layer determine a certain thickness of the Pt layer for the STT efficiency.
For the Ta/Pt(t)/FM structure in this work, we optimized the Pt thickness to 1.8 nm to have a maximum STT efficiency.
This work highlights the spin diffusion process in the current induced STT and may be helpful in designing STT-based devices.

\section{ACKNOWLEDGMENTS}
This work is supported by the National Natural Science Foundation of China (NSFC No. 11774200) grants and the Shandong Provincial Natural Science Foundation (No. ZR2019JQ02).

\end{document}